\documentstyle[12pt]{article}
\long\def\@makefntext#1{\parindent 0cm\noindent
\hbox to 1em{\hss$^{\@thefnmark}$}#1}

\begin{document}

\begin{center}
{\Large\bf
 Reply to the comment by Park and Ho}\\
\vspace{.4in}
{S.~C{\sc arlip}\footnote{\it email: carlip@dirac.ucdavis.edu}\\
       {\small\it Department of Physics}\\
       {\small\it University of California}\\
       {\small\it Davis, CA 95616}\\{\small\it USA}}
\end{center}

\addtocounter{footnote}{-1}

Park and Ho \cite{Park}  correctly point out an error in the surface term (9) 
of Ref.\ \cite{Carlip1}, and the consequent lack of antisymmetry in the 
Poisson brackets (13) of that paper.  The root of the problem is that the 
surface term fails to make the action sufficiently ``differentiable'': the 
diffeomorphisms of interest, given by eqns.\ (17)-(18) of Ref.\ \cite{Carlip1}, 
do not satisfy the boundary conditions (described after eqn.\ (10)) needed 
in order for the total surface term in the variation of the action to vanish.  
An interesting alternative has been proposed by Sovoliev \cite{Sovoliev}, who 
suggests an approach to boundary terms that does not require ``differentiability'' 
of the action, but it is not clear to me whether this completely resolves the 
problem. 

Despite this error, however, the overall conclusion of Ref.\ \cite{Carlip1} 
remains valid.  The relevant surface terms have been analyzed more thoroughly, 
in a manifestly covariant manner, in Ref.\ \cite{Carlip2}, where it is shown 
that the Virasoro algebra of Ref.\ \cite{Carlip1} is reproduced with the correct
values of central charge and $L_0$ needed to explain black hole entropy.   The
details of the relation between the covariant phase space approach of Ref.\
\cite{Carlip2} and the canonical approach of Ref.\ \cite{Carlip1} have not yet
been worked out, however.  Moreover, as Ref.\ \cite{Carlip2} demonstrates, 
there are still unanswered questions concerning the proper generic boundary 
conditions for a black hole horizon, so the subject cannot be considered closed.

While the the error pointed out by Park and Ho is a real one, their analysis
is slightly misleading in one respect.  Their criticism (b) implicitly assumes
that because a nonrotating black hole has no {\em preferred\/} angular
direction, diffeomorphisms like those of eqn.\ (18) of Ref.\ \cite{Carlip1}
should be taken to be independent of angular coordinates.  But as observed
in Ref.\ \cite{Carlip2}, the generator of diffeomorphisms (with the correct 
boundary terms) exists only when an angular dependence is included.  For
a static black hole, this angular dependence is nearly arbitrary---different
choices lead to isomorphic Virasoro algebras---but it cannot be neglected.  
As a result, criticism (b) is not valid.

The Comment by Park and Ho also raises an interesting issue that deserves
to be investigated further.  Their ``anomalous boundary contribution'' to the 
transformation of, for example,  $g_{rr}$ is determined by looking at the 
variation $\delta L[\xi]$ of the generator of diffeomorphisms and extracting 
the coefficient of $\delta\pi^{rr}$.  This certainly seems to be a valid procedure.
On the other hand, however, the relevant variation $\delta L[\xi]$ is
\begin{equation}
\delta L[\xi] \sim \int_{r=r_+} {1\over f}n_r{\hat\xi}^r g_{rr}\delta\pi^{rr},
\quad
\hbox{with}\ 
\pi^{rr} = -{1\over f}\sqrt{\sigma}\sigma^{\alpha\beta}K_{\alpha\beta}.
\label{a1}
\end{equation}
If one uses the boundary conditions (3) and (4) of Ref.\ \cite{Carlip1}, one 
finds that this variation vanishes.  Park and Ho obtain a nonzero contribution 
to $\delta g_{rr}$ because they divide $\delta L[\xi]$, which vanishes at $r=r_+$,
by $\delta\pi^{rr}$, which also vanishes.  

Now, in the standard approach of Regge and Teitelboim \cite{Regge} to surface
terms and ADM mass, one computes the variation $\delta L[\xi]$, takes an
appropriate limit to go to a boundary such as spatial infinity, and then uses
the limiting variation to determine a boundary term.  What Park and Ho
have demonstrated is that this process does not necessarily ``commute'' with 
the process of functionally differentiating to obtain Poisson brackets.  It may
be that this paradox can be resolved by correctly incorporating boundary
conditions into the definition of Dirac brackets, but further investigation
seems warranted.

\vspace{1.5ex}
\begin{flushleft}
\large\bf Acknowledgements
\end{flushleft}

This work was supported in part by Department of Energy grant 
DE-FG03-91ER40674.

\end{document}